\documentstyle[11pt,newpasp,twoside,epsf]{article}
\def\edcomment#1{\iffalse\marginpar{\raggedright\sl#1\/}\else\relax\fi}
\reversemarginpar

\begin{document}
\title{Atmosphere-Corrected Phase-Referencing}
\author{Andreas Brunthaler}
\affil{Max-Planck-Institut f\"ur Radioastronomie, Auf dem H\"ugel 69, 53121 Bonn, Germany}
\author{Mark J. Reid}
\affil{ Harvard-Smithsonian Center for Astrophysics, 60 Garden Street, MS~42, Cambridge, MA 02138, USA }
\author{Heino Falcke}
\affil{ASTRON, P.O. Box 2, 7990 AA Dwingeloo, The Netherlands}
\affil{Max-Planck-Institut f\"ur Radioastronomie, Auf dem H\"ugel 69, 53121 Bonn, Germany}

\begin{abstract}
One major problem of phase-referencing VLBI observations are phase errors due to the unknown tropospheric zenith delay at each antenna. These errors degrade the quality of the phase-referenced image and limit the achievable astrometric accuracy. We will present and compare two independent methods to estimate the zenith delay offset at each antenna. The zenith delay offsets can then be used to correct the data. These corrections improve the quality of the phase-referenced image and an astrometric accuracy of 10 $\mu$as can be achieved. With this accuracy, measurements of proper motions in the Local Group become feasible.
\end{abstract}

\section{Introduction}
Phase-referencing has become a standard technique in Very Long Baseline Interferometry (VLBI) to measure accurate relative positions of radio sources. The concept of phase-referencing is based on the
assumption that the phase errors of two sources with a small angular
separation on the sky are similar. One observes a target source
between two scans on a calibrator. Then one can interpolate the calibration
between the two scans and apply this interpolated phase corrections to the
phase of the target source. The phase-referenced difference phase contains information only about the target source structure and its position relative to the calibrator and contains noise from interpolation errors. 

One major problem of phase-referencing VLBI observations is the unknown tropospheric zenith delay at each antenna. The VLBA correlator model uses a seasonally averaged and latitude-dependent atmospheric model~(Niell 1996), which can misestimate the zenith delay by a few centimeters.
The excess path length caused by the dry troposphere to a radio signal from a source with zenith angle $Z$ is given by
\begin{eqnarray}
l\approx0.228\frac{\mathrm{cm}}{\mathrm{mb}}P_0\,\mathrm{sec}Z(1-0.0013\,\mathrm{tan}^2Z)
\end{eqnarray}
where $P_0$ is the total pressure at the surface (see equation 13.41 in Thompson, Moran, \& Swenson 2001). This excess path length will be different for the calibrator and the target source since both sources usually have different zenith angles.  Hence, a residual error will remain after phase-referencing. These residual errors degrade the quality of the phase-referenced image and limit the achievable astrometric accuracy (e.g., Beasley \& Conway 1995). If the correlator model misestimates the zenith delay $l_0$ by 3 cm ($\approx 0.1$ ns), one would expect after phase-referencing delay errors of 0.2 cm for zenith angles of $\approx 60^\circ$ and source separations of $\approx 1^\circ$. However, these errors can be corrected if the true zenith delay at each antenna is known.

\section{Atmosphere-Corrected Phase-Referencing}
\subsection{Phase fitting}
The fringe-phase of the target source after phase-referencing is mainly a sum of the position offset and the effects of the zenith delay error if the source structure is negligible. The phase induced by the position offset shows a simple 24 hour cosine behavior, while the phase errors from the zenith delay error follow a more complex dependence on the zenith angle according to equation 1. Because of the different behavior of the two contributions, it is possible to separate both effects and to estimate the position offset as well as the zenith delay error.

One can compute a model phase with a position offset (two parameters) and a zenith delay error at each antenna (one parameter for each antenna) and perform a least-squares fit to the phase data. This method has been applied successfully by Reid et al. (1999) to measure the proper motion of Sgr A$^*$. In Brunthaler et al. (2002) we applied this method to phase-referencing observations of extragalactic H$_2$O masers in IC\,10 with respect to the background quasar J0027+5958 and achieved an astrometric accuracy of $\approx 10~\mu$as.

\begin{table}
\caption{Positions offsets of IC\,10 in right ascension ($\Delta\alpha$) and declination ($\Delta\delta$) relative to the phase center from the model fit to the phase data (A), the corrected map using the phase data (B) and the corrected map using the geodetic-like observations (C).}
\begin{tabular}{|c|cc|cc|cc|}
\hline
&\multicolumn{2}{c|}{A}&\multicolumn{2}{c|}{B}&\multicolumn{2}{c|}{C}\\
\hline
Date& $\Delta\alpha$~[$\mu$as]& $\Delta\delta$~[$\mu$as]& $\Delta\alpha$~[$\mu$as]& $\Delta\delta$~[$\mu$as]& $\Delta\alpha$~[$\mu$as]& $\Delta\delta$~[$\mu$as]\\
\hline
2002/01/12&-4&112&-1&103&13&101\\
2002/01/17&-11&134&-8&125&1&135\\
\hline
Mean&-8$\pm$4&123$\pm$11&-5$\pm$4&114$\pm$11&7$\pm$6&118$\pm$17\\
\hline
\end{tabular}
\end{table}

Second epoch observations of the H$_2$O masers in IC\,10 were made in January 2002 and two observations were made within five days to check the accuracy and repeatability of our results. The data were calibrated with standard techniques using the AIPS software package. A priori amplitude calibration was applied using system temperature measurements and standard gain curves. A fringe fit was performed on J0027+5958 and the solutions were applied to IC\,10.

Then we modeled the phase data of IC\,10 as described above. The position offset for the two observations are given in Table~1~(A). The positions obtained from the two observations are consistent and have an rms of $\approx 10~\mu$as. The zenith delay offsets are typically of the order of a few centimeters.  The values of the first observation on 2002 January 12 are given in the second column of Table~2. 

\subsection{Phase correction}

\begin{figure}[!h]
\plotone{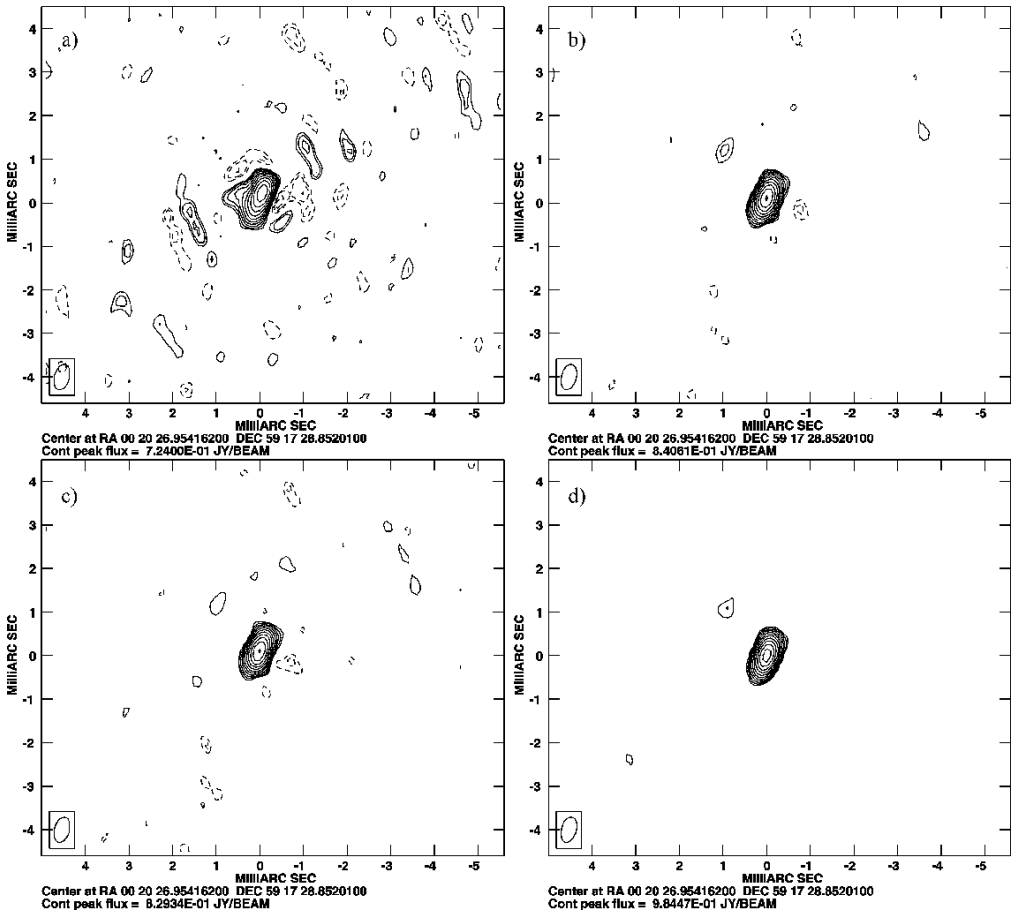}
\caption{Phased-referenced image of a H$_2$O maser spot in IC\,10 without correction a), with atmosphere correction from the fit to the phases b), with corrections form the geodetic-like observations c) and after phase self-calibration d). The contour levels start at 25 mJy and increase by a factor of $\sqrt{2}$ in all images. The peak-to-noise ratios are 77, 101, 100 and 151 respectively.}
\end{figure}

The fitted zenith delay offsets can be used to correct the {\it uv}-data itself. This correction was done using a new version of the AIPS task CLCOR with OPCODE='ATMO'. Fig.~1 shows the phase-referenced images of a strong H$_2$O maser component in IC\,10  with (b) and without (a) the corrections. The noise in the image is reduced while the peak flux increased in the corrected image. The peak-to-noise ratio in the corrected image is 101 compared to 77 in the uncorrected image. The positions of the maser feature in the corrected maps for the two observations are given in Table~1~(B) and agree within 10 $\mu$as with the positions derived from the fit to the phases. The positions obtained from the fit to the phases and from the corrected maps are not independent and the agreement is not surprising. However, it is a consistency check for the method.

\subsection{Geodetic-like observations}
A second, independent method to estimate the zenith delay offset is to use {\it geodetic-like} observations. Here, a number of bright quasars with positions known to better than 1 mas were observed for 45 minutes before and after the phase-referencing observation. The quasars were observed at different elevations with a frequency setup of 8 IFs spread over 450 MHz. A fringe-fit on the quasars yielded then a multi-band delay and a rate for each baseline and source. These multi-band delays and rates were then fit with a model that consisted of a zenith delay offset and drift at all antennas as well as a clock offset and drift at all antennas except the reference antenna. The parameters of the best model-fit are given in Table~2. The differences between the observed multi-band delay and the model are less than 0.1 ns. 

The zenith delay offsets obtained from the fit to the phases and from the geodetic-like observations are in general comparable although they can be different for some antennas. The differences between the two methods are not surprising, since the model in the geodetic-like observations includes not only a zenith delay offset but also a drift, a clock offset and a clock drift.

\begin{table}
\caption{Fit parameters for the observation on 2002 January 12: Zenith delay offset ($\tau_{0,p}$) from the fit to the phase data , zenith delay offset ($\tau_0$) and rate ($\dot\tau_0$) and clock offset ($\Delta t$) and rate ($\dot{\Delta t}$) from the geodetic-like observations. The errors are the formal errors of the least-squares fit.}
\begin{tabular}{|c|c|cccc|}
\hline
Antenna&$\tau_{0,p}~[\mathrm{cm}]$&$\tau_0~[\mathrm{cm}]$&$\dot\tau_0~[\mathrm{cm~h}^{-1}]$&$\Delta t [\mathrm{ns}]$&$\dot{\Delta t} [\mathrm{ns~d}^{-1}]$\\
\hline
BR &2.5$\pm$0.6  & 1.6$\pm$0.1   & 0.4$\pm$0.03 &-0.24$\pm$0.01&-1.52$\pm$0.03  \\
FD &4.4$\pm$0.3  & 1.7$\pm$0.05  & 0.3$\pm$0.01 &-0.08$\pm$0.01 &-0.84$\pm$0.02  \\
HN &7.0$\pm$0.2  & 6.4$\pm$0.1   & 0.3$\pm$0.03 & 0.37$\pm$0.01 & 0.47$\pm$0.04  \\
KP &1.0$\pm$0.3  &-0.1$\pm$0.05  & 0.3$\pm$0.01 & --   & --    \\
LA &-0.4$\pm$0.4 &-1.7$\pm$0.07  & 0.3$\pm$0.02 &-0.12$\pm$0.01 &-0.17$\pm$0.02  \\
MK &-3.3$\pm$0.1 &-4.0$\pm$0.06  & 0.3$\pm$0.01 &-0.93$\pm$0.01 &-0.97$\pm$0.03  \\
NL &9.4$\pm$0.4  & 7.8$\pm$0.1   & 0.2$\pm$0.02 & 0.42$\pm$0.01 & 0.51$\pm$0.03  \\
OV &4.7$\pm$0.4  & 2.2$\pm$0.07  & 0.3$\pm$0.01 & 0.00$\pm$0.01 &-0.62$\pm$0.02  \\
PT &-0.2$\pm$0.4 &-1.8$\pm$0.07  & 0.2$\pm$0.01 & 0.08$\pm$0.01 & 1.05$\pm$0.02  \\
\hline
\end{tabular}
\end{table}

The fit values of the zenith delay and clock offsets can also be used to correct the {\it uv}-data. A corrected image is seen in Figure~1~(c) and the peak-to-noise ratio is 100. The positions of the maser feature in the corrected maps are given in Table~1~(C). They are also consistent with the positions of the fit to the phases.

\section{Conclusions}
We have presented a technique to correct phase-referencing observations for inaccurate zenith delays in the VLBA correlator model. The quality of the atmosphere-corrected phase-referenced images can be improved by this technique. The peak-to-noise ratio in the corrected images increases by $\approx 30~\%$. By using this technique, an astrometric accuracy of 10 $\mu$as can be achieved. 

Two independent methods were presented to estimate the zenith delay offsets at each antenna. The images obtained from the correction with the zenith delay offsets from the fit to the phases (Fig.1~(b)) and from the geodetic-like observations (Fig.1~(c)) are very similar and the positions measured from these images are consistent. This is a proof of concept for our technique.

The astrometric accuracy of this technique enables us to detect significant proper motions of galaxies in the Local Group out to a distance of 800 kpc within a few years. One needs at least three epochs of observations to make reliable statements about measured proper motions. Further observations are underway and the results will be presented elsewhere.

%\begin{figure}[!h]
%\plotone{multiplot.eps}
%\caption{Multi-band delay for three baselines. Shown is the data (crosses), model (squares) and residuals (circles).}
%\end{figure}

\end{document}